\documentclass[prb,twocolumn,floatfix,notitlepage,superscriptaddress,longtable]{revtex4-2}
\usepackage{amsmath}
\usepackage{lipsum} % for some text
\usepackage{verbatim}
\usepackage{epsfig}
\usepackage{subfigure}
\usepackage{graphicx}
\usepackage{amsfonts}
\usepackage[figuresright]{rotating}
\usepackage{amssymb}
\usepackage{amsmath}
\usepackage{psfrag}
\usepackage{esint} %double integrals
\usepackage{bm}% bold math
\usepackage[colorlinks,linkcolor=blue,anchorcolor=blue,citecolor=blue,urlcolor=blue]{hyperref}
\usepackage[version=4]{mhchem}
\usepackage[svgnames]{xcolor}
\def\be{\begin{equation}} \def\ee{\end{equation}}
\def\bea{\begin{eqnarray}} \def\eea{\end{eqnarray}}

\def\k{{\bf k}}

\renewcommand{\vec}[1]{\mathbf{#1}}

\newcommand{\ket}[1]{| #1 \rangle}
\newcommand{\bra}[1]{\langle #1 |}

\def\bpm{\begin{pmatrix}} \def\epm{\end{pmatrix}}

\DeclareMathOperator{\Tr}{Tr}

\definecolor{Yangcolor}{RGB}{1, 148, 109}

\makeatletter
\newcommand*{\balancecolsandclearpage}{%
  \close@column@grid
  \clearpage
}
\makeatother

\begin{document}

\title{Universal nonadiabatic energy pumping in a quasiperiodically driven extended system}

\author{Zihao Qi}
\affiliation{Institute of Quantum Information and Matter and Department of Physics,California Institute of Technology, Pasadena, CA 91125, USA}
\author{Gil Refael}
\affiliation{Institute of Quantum Information and Matter and Department of Physics,California Institute of Technology, Pasadena, CA 91125, USA}
\author{Yang Peng}
\email[Corresponding author: ]{yang.peng@csun.edu}
\affiliation{
Department of Physics and Astronomy, California State University, Northridge, Northridge, California 91330, USA}
\affiliation{Institute of Quantum Information and Matter and Department of Physics,California Institute of Technology, Pasadena, CA 91125, USA}
\date{\today}

\begin{abstract}
The paradigm of Floquet engineering of topological states of matter can be generalized into the time-quasiperiodic scenario, where a lower dimensional time-dependent system maps into a higher dimensional one by combining the physical dimensions with additional synthetic dimensions generated by multiple incommensurate driving frequencies. Different than most previous works in which gapped topological phases were considered, we propose an experimentally realizable, one dimensional chain driven by two frequencies, which maps into a gapless Weyl semimetal in synthetic dimension. Based on analytical reasoning and numerical simulations, we found the nonadiabatic quantum dynamics of this system exhibit energy pumping behaviors characterized by universal functions. We also numerically found such behaviors are robust against a considerable amount of spatial disorder. 
\end{abstract}

\maketitle

\section{\label{sec:level1}Introduction}
The idea that topological phases of matter out of equilibrium can be engineered with external time-dependent drives has been thoroughly investigated over the past few years in the Floquet paradigm~\cite{Oka2009,Inoue2010,Kitagawa2010,Kitagawa2011,Lindner2011,Rudner2013,Lindner2013,Titum2016,Khemani2016,Keyserlingk2016c,Rahul2016,Potter2016,Potirniche2017,Else2016,Else2017,Yao2017,Zhang2017,Choi2017, DrivenSYK}, and was only recently generalized to the time-quasiperiodic realm, where a quantum system is driven by external drives at multiple mutually incommensurate frequencies, under the framework pioneered by Ref.~\cite{GilPRX}. The key insight of that work is that each drive at a particular frequency maps into a synthetic dimension of energy stored in that drive, which is quantized in unit of its frequency.
In fact, the concept of synthetic dimensions has recently emerged as a powerful way to emulate topological phases of matter, which are now of great interest across many areas of physics~\cite{Ozawa2019}.
Among various approaches to engineer synthetic dimensions, the idea based on quasiperiodic drives was pursued and
generalized by several theorists~\cite{YangPRB,PhysRevB.99.064306,Nathan2019,crowley2020,Nathan2021,PhysRevResearch.2.043411,Psaroudaki2021,PhysRevLett.126.106805}, as well as realized in experiments~\cite{PhysRevLett.125.160505,PhysRevLett.126.163602}.

In general, when a $d$-dimensional system is subject to $n$ quasiperiodic drives with mutually incommensurate frequencies, quantum states are dressed by all harmonics of the driving frequencies. The amplitudes for the harmonics form new degrees of freedom, thereby effectively raising the dimensionality of the system from $d$ to $d+n$. Furthermore, different driving frequencies, if collected into a vector, resemble a homogeneous electric field operating in the synthetic space. In an extended system, when the external drives are not homogeneous in the physical dimensions, a synthetic magnetic field can also be realized~\cite{YangPRB}. 

\begin{figure}
    \centering
    \includegraphics[width=0.49\textwidth]{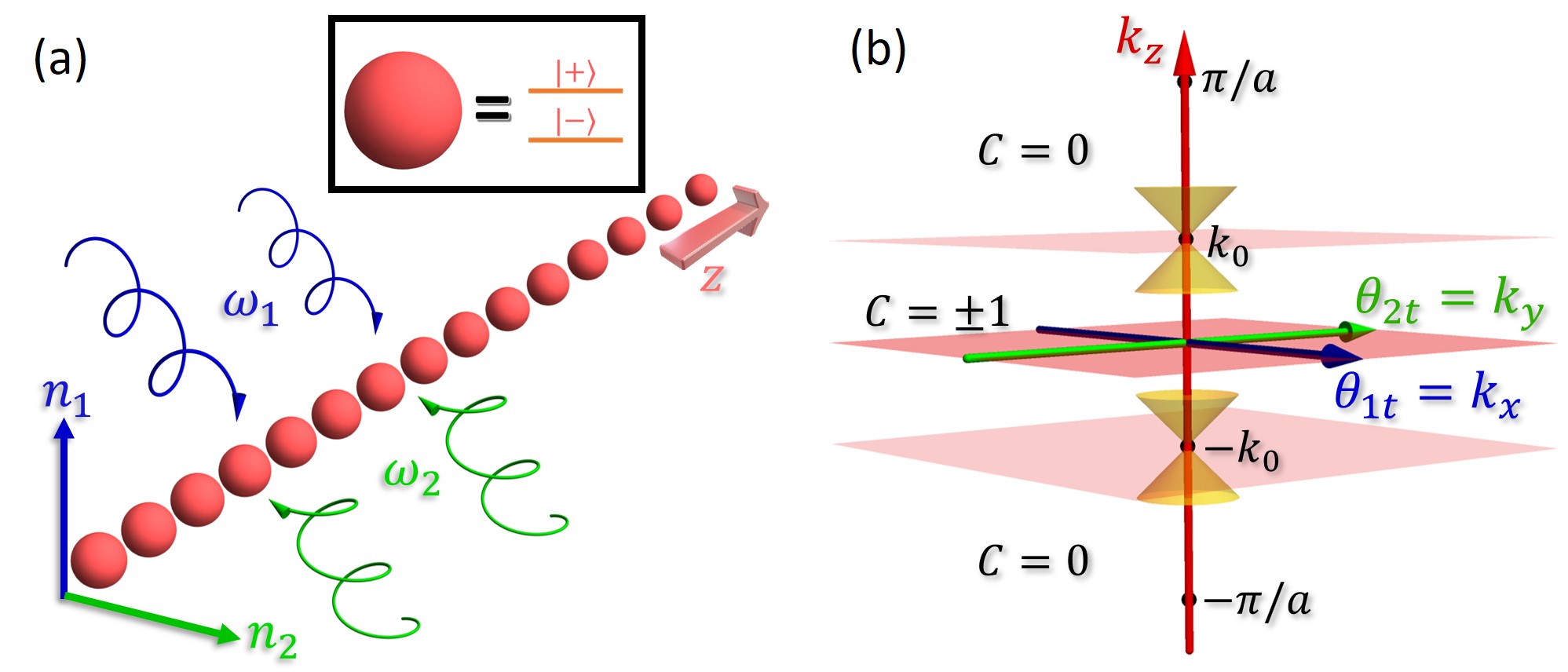}
    \caption{Model illustrations. (a) A chain aligned along $z$ direction is quasiperiodically driven by two incommensurate frequencies $\omega_1$ and $\omega_2$, which map into synthetic dimensions of energy quanta in units of these frequencies, counted by $n_1$ and $n_2$. Inset: each site should be effectively a two-level system. (b) Apart from having a synthetic electric field, the driven chain maps to a synthetic WSM, where the Brillouin zone is parametrized by $\theta_{jt}=\omega_{j}t+\theta_{j0}$ with $j=1,2$ and Bloch momentum $k_z$, when periodic boundary condition along $z$ direction is assumed. The locations of the Weyl nodes are indicated at $\pm k_0$, and the linear dispersion is shown along $k_z$. The model can be regarded as 2D systems characterized by the Chern number $C$ at each $k_z$ except for at $k_z=\pm k_0$, where the 2D system is gapless and $C$ is undefined, as indicated by red planes. The Chern number $C=\pm 1$ when $-k_0<k_z<k_0$, and $C=0$ elsewhere.}
    \label{fig:model_illustration}
\end{figure}

In response to the synthetic electromagnetic fields, the system generates currents flowing along the synthetic dimensions, which can be interpreted as the energy current pumped into each drive. In the adiabatic regime, i.e., when these driving frequencies are much smaller than the energy gap of the $(d+n)$-dimensional synthetic lattice system, linear response theory can be applied.
Based on this, Ref.~\cite{GilPRX} considered a single spin-1/2 under two incommensurate drives and demonstrated the quantum-Hall-like topological energy pumping phenomenon, since such a system can be mapped into a synthetic 2D Chern insulator.
Ref.~\cite{PhysRevB.99.064306} further generalized this by going beyond the adiabatic limit and showed that quantized energy pumping is a dynamical signature of quasienergy states in the topological class of dynamics. 

So far, synthetic-space topological phases considered were exclusively gapped (such as the Chern insulator). In contrast, Ref.~\cite{crowley2020} considered the same driven spin-1/2 model, but allowed the mapped synthetic 2D lattice system to become gapless at the phase transition point between a topologically trivial insulator and a nontrivial Chern insulator. 
Despite the nonadiabatic nature of this gapless system, it was shown that the energy pumping power can be characterized by some universal scaling forms, and the short-time topological energy pumping becomes half-integer quantized at the gapless point. This result was even demonstrated experimentally~\cite{PhysRevLett.125.160505}.

From a practical perspective, the quasiperiodically driven quantum systems exhibiting robust/universal energy pumping behaviors are potentially crucial for quantum technological applications, especially for quantum communications and quantum computing purposes, as the energy flow between different subsystems can be well controlled. Understanding the quantum dynamics beyond the adiabatic limit is important because it greatly expands the parameter regime where the energy-pumping device operates. On the other hand, an extended-system-based device is favorable, as its energy pumping power can be manipulated by changing the system size~\cite{YangPRB}, which adds more controllability.

Using this as a motivation, we propose a quasiperiodically driven 1D system
which maps into a synthetic gapless topological phase, in particular a Weyl Semimetal (WSM).
We show that a synthetic WSM can be realized by driving a 1D extended chain with two external drives at incommensurate frequnecies, as shown in Fig.~\ref{fig:model_illustration}(a). 
Despite the system's intrinsic gapless and hence nonadiabatic nature, we find both analytically and numerically that this system exhibits robust (disorder insensitive) energy pumping behaviors characterized by universal scaling functions. 

The rest of the paper is organized as follows. In Sec.~\ref{sec:review}, we review the mapping between a $d$-dimensional system under $n$ drives and a $d+n$ dimensional system, as well as energy pumping in such driven systems.  In Sec.~\ref{sec:pumpingpower}, we introduce the model of a 1D chain driven by two incommensurate frequencies, and show that it maps to a synthetic WSM (illustrated in Fig.~\ref{fig:model_illustration}(b)). Moreover, we analytically derive the universal functional form characterizing the energy pumping behavior in this system. In Sec.~\ref{sec:numerical}, we provide numerical support for our analytical results. We also discuss the effects of the boundary condition, as well as spatial disorder. In particular, we find that our results are still valid even when the disorder strength is comparable to the overall energy scale of the model. In Sec.~\ref{sec:discussions}, we further discuss the timescale at which spins decouple from the magnetic field. From this, we derive the expected energy pumping behavior at intermediate/long times. Finally, we conclude our work in Sec.~\ref{sec:conclusions}, where we also comment on the possible interaction effects on our noninteracting system.
%\section{Review of previous work}

\section{Time-quasiperiodic quantum systems \label{sec:review}}
\subsection{\label{sec:quasiperiodic}Synthetic dimensions from quasiperiodic drives}
In this section, we follow Ref. \cite{GilPRX} and review the mapping between a $d$-dimensional system under $n$ mutually incommensurate drives and a ($d+n$)-dimensional system. Note that any two of the driving frequencies must not be commensurate, otherwise they effectively act as one drive with a longer period.

Consider a non-interacting Hamiltonian defined on a lattice in $d$ dimensions. In the second quantization formalism, it can be written as:
\begin{equation}
    H = \sum_{\vec{x},\vec{x'}} \Psi^\dagger(\vec{x}) \mathcal{H}(\vec{x},\vec{x'}) \Psi(\vec{x'})
\label{eq:Hamiltonian}
\end{equation}
where $\Psi^\dagger(\vec{x})$ ($\Psi(\vec{x})$) is the creation (annihilation) operator at position $\vec{x}$, and the double sum runs over all lattice sites.

Introducing $n$ $2\pi$-periodic variables
of the form $\theta_{it}(\vec{x}) = \omega_i t + \theta_{i0}(\vec{x})$ to the on-site Hamiltonian, 
namely $\mathcal{H}(\vec{x},\vec{x}) \to \mathcal{H}(\vec{x},\vec{x};\boldsymbol{\theta_{t}}(\vec{x}))$, with $\boldsymbol{\theta_{t}}(\vec{x})=(\theta_{1t}(\vec{x}),\dots, \theta_{nt}(\vec{x}))$, and $\omega_i$ corresponding to the frequency of the $i$th external drive, we can write down the general wave function as
\begin{equation}
    \ket{\psi(t)} = e^{-iEt} \sum_{\vec{x}, \vec{m}} \Phi_{\vec{m}}(\vec{x}) e^{-i \vec{m} \cdot \text{\boldmath$\omega$} t} \ket{\vec{x}},
\label{eq:wf}
\end{equation}
where $E$ is the quasi-energy of the state,  $\vec{m} = (m_1, m_2, ... m_n) \in \mathbb{Z}^n$, and $\text{\boldmath$\omega$} = (\omega_1, \omega_2, ..., \omega_n)$ is the vector of all $n$ driving frequencies.

Plugging the above ansatz into the Schr\"{o}dinger equation,
we obtain
\begin{align}
     &(E+\vec{m} \cdot \text{\boldmath$\omega$}) \Phi_{\vec{m}}(\vec{x})  =  \sum_{\vec{x'} \neq \vec{x}} \mathcal{H}(\vec{x},\vec{x'})
    \Phi_{\vec{m}}(\vec{x'}) +  \nonumber
    \\
    &\sum_{\vec{q}} H_{\vec{q}}(\vec{x}) e^{- i \vec{q} \cdot \boldsymbol{\theta_{i0}}(\vec{x})} \Phi_{\vec{m-q}}(\vec{x'}),
\label{eq: tightbindingeq}
\end{align}
where $H_{\vec{q}}(\vec{x})$ is defined via the Fourier decomposition
\begin{equation}
    \mathcal{H}(\vec{x},\vec{x};\boldsymbol{\theta_t}(\vec{x})) = \sum_{\vec{q}} H_{\vec{q}}(\vec{x})e^{-i \vec{q}\cdot\boldsymbol{\theta_t}(\vec{x})},
\end{equation}
and we also used the following fact about the Floquet modes: $e^{-i \vec{q} \cdot \text{\boldmath$\omega$} t} \Phi_{\vec{m}} = \Phi_{\vec{m-q}}$. Note that Eq.~\ref{eq: tightbindingeq} is the same eigenvalue equation describing an electron hopping model on a $d+n$-dimensional lattice under a scalar potential $A_0(\vec{x}, \vec{n}) = -\vec{n} \cdot \text{\boldmath$\omega$}$ and a vector potential $\vec{A}(\vec{x}, \vec{n}) = (\vec{0}_d, \text{\boldmath$\phi$(x)})$, where $\vec{0}_d$ is the $d$-dimensional zero vector \cite{YangPRB}.
Equivalently, we have obtained the synthetic electromagnetic fields emerging from the drives, given by $\vec{E} =(\boldsymbol{0},\boldsymbol{\omega})$, and $\vec{B}$ with component $B_k = \sum_{ij}\epsilon_{ijk}\partial\phi_j(x)/\partial x_i$,
where the index $i$ is restricted to only physical-dimensional directions, whereas $j$ is restricted to only synthetic-dimension directions. 

\subsection{Energy pumping in a quasiperiodically driven system\label{sec:density_matrix}}
One peculiar feature of quasiperiodically driven systems is energy pumping: namely, energy currents can flow between 
different external drives. The energy pumping phenomena of the system can be described using the following framework. 

Consider a time-quasiperiodic Hamiltonian that depends on time $t$ through $n$ $2\pi$-periodic variables: $H(t)\equiv H(\theta_{1t}, \theta_{2t},\dots, \theta_{nt})\equiv H(\boldsymbol{\theta_t})$, where $\theta_{it}=\omega_i t + \theta_{i0}$. Let us assume that the system consists of $N$ noninteracting electrons, and that they are initially (at $t=0$) in equilibrium at zero temperature. Such a system can be described by the density matrix 
\begin{equation}
    \hat{\rho}(t) = \sum_{j=1}^N \ket{\psi_j(t)}\bra{\psi_j(t)},
    \label{eq:dm_N}
\end{equation}
where $\ket{\psi_j(0)}$ is the $j$th eigenstate, sorted by eigenvalues from low to high, of the Hamiltonian $H$ at time $t=0$.
Note that the $\{\ket{\psi_j(t)}\}$ are not instantaneous eigenstates of $H(t)$ at finite time $t$. The evolution of an individual state $\ket{\psi_j(t)}$ is described by the Schr\"odinger equation
\begin{equation}
    i\frac{d}{dt} \ket{\psi_j(t)} = H(t)\ket{\psi_j(t)},
\end{equation}
or equivalently, the density matrix satisfies the Liouville–von Neumann equation
\begin{equation}
    i\frac{d}{dt}\hat{\rho}(t) = [H(t),\hat{\rho}(t)].
    \label{eq:dm_evolution}
\end{equation}

Since the Hamiltonian is explicitly time-dependent, the total energy of the system is not conserved. Instead, the expectation value of the total energy can be expressed as:
\begin{equation}
    E_{tot} (t) = \Tr\left[\hat{\rho}(t)H(t)\right],
\end{equation}
where ``Tr'' denotes the operator/matrix trace. By taking the time derivative of $E_{tot}(t)$, one can also define the rate of the total energy pumped into the system as
\begin{equation}
    P_{tot}(t) = \Tr\left[\left(\frac{d}{dt}\hat{\rho}(t)\right)H(t)\right] + \Tr\left[\hat{\rho}(t)\left(\frac{d}{dt}H(t)\right)\right].
\end{equation}

In the above equation, the first term vanishes because of Eq.~(\ref{eq:dm_evolution}) and the cyclic property of the trace. If we further use
\begin{equation}
    \frac{d H(t)}{dt} = \sum_{l=1}^n \omega_l \frac{\partial}{\partial \theta_{lt}} H(\theta_{1t},\theta_{2t},\dots,\theta_{nt}),
\end{equation}
we can express the total energy pumping power as
\begin{align}
    P_{tot}(t) &= \sum_{l=1}^n \omega_l \Tr\left[\hat{\rho}(t) \frac{\partial H}{\partial \theta_{lt}}\right] \nonumber \\
    &= \sum_{l=1}^n \omega_l\sum_{j=1}^N \left \langle \psi_j(t) \middle| \frac{\partial H}{\partial \theta_{lt}} \middle| \psi_j(t) \right \rangle,
\end{align}
where each term in the first sum can be interpreted as the energy pumping power provided by an individual drive (at a given frequency $\omega_l$).

Note that if we interpret $H(\boldsymbol{\theta_{t}})$ as a Bloch Hamiltonian, where $\boldsymbol{\theta_t}$ is defined in the $n$-dimensional synthetic Brillouin zone, we can regard $\hat{v}_l = \partial H/\partial \theta_{lt}$ in the above equation as the velocity operator in the $l$th direction on the synthetic Floquet lattice. 

If the single-particle gap between the occupied and unoccupied states is much larger than the driving frequency $\omega_l$, we can apply the adiabatic transport theory \cite{RevModPhys.82.1959}
to obtain the energy current that arises due to the $l$th drive:
\begin{align}
    J_l &= \omega_l\Tr\left[\hat{\rho}(t)\frac{\partial H}{\partial \theta_{lt}}\right] \nonumber \\
    &\simeq \omega_l\sum_{j=1}^{N}\left\{ \frac{\partial E_l(\boldsymbol{\theta}_t)}{\partial \theta_{lt}} - \sum_{k\neq l}\omega_k \Omega_{lk}^{(j)}\right\},
    \label{eq:adiabatic_current}
\end{align}
where $E_l(\boldsymbol{\theta}_t)$ is the $l$th instantaneous eigenvalue at 
$\boldsymbol{\theta}_t = (\theta_{1t},\theta_{2t},\dots)$, and 
\begin{equation}
    \Omega_{lk}^{(j)}(\boldsymbol{\theta}_t) = i\left[
    \left\langle \frac{\partial \chi_j(\boldsymbol{\theta}_t)}{\partial \theta_{lt}} \middle|
    {\frac{\partial \chi_j(\boldsymbol{\theta}_t)}{\partial \theta_{kt}}} \right \rangle -
    (l \leftrightarrow k)
    \right]
    \label{eq:berrycurvature}
\end{equation}
is the Berry curvature tensor for the instantaneous eigenstate $\chi_j(\boldsymbol{\theta}_t)$. When averaging over time (equivalently, over $\boldsymbol{\theta}_t$), the first term in Eq.~(\ref{eq:adiabatic_current}) vanishes as a consequence of Bloch oscillations, whereas the second term produces a robust topological energy pumping rate proportional to the Chern number by sampling $\Omega_{lk}^{(j)}$ over the two-dimensional closed manifold spanned by $(\theta_{lt},\theta_{kt})$.

On the other hand, however, when there is no energy gap, or when the gap is comparable to the driving frequencies, the adiabatic transport theory will not be applicable. In this case, excited states will be populated due to nonadiabatic transitions, and as a result, the robust topological energy pumping phenomena will not be expected.

\section{Energy pumping powers in a synthetic Weyl semimetal \label{sec:pumpingpower}}
In this section, we introduce a chain of two-level systems under two external drives with mutually incommensurate frequencies. We show that such a time-quasiperiodic system maps to a 3D Weyl semimetal (WSM), under the approach outlined in Sec.~\ref{sec:quasiperiodic}. 

Because the system hosts synthetic gapless points, i.e. the synthetic Weyl nodes, it is expected that the adiabatic condition can never be satisfied in the entire Brillouin zone (BZ).
However, as we will show below, to a large extent the energy pumping can still be analytically characterized in terms of two universal functions corresponding to the topological and the excitation components of the energy pumping.

In the following, before introducing the 1D driven chain, we first review a simple model in which non-adiabatic processes appear, following Ref. \cite{crowley2020}. 
Understanding the properties of such a simpler model will be helpful to 
the discussion of our more complicated extended system.

\subsection{\label{sec:spin-half}Driven spin-half near a gapless point}
Consider a single spin subjected to two drives with incommensurate frequencies, described by the following Hamiltonian
\begin{equation}
    H(t) = -\frac{1}{2} \boldsymbol{B}(t) \cdot \boldsymbol{\sigma} \\
\label{eq:driven1/2},
\end{equation}
where the magnetic field is quasiperiodic in time:
\begin{equation}
    \boldsymbol{B}(t) = B_0 (\sin\theta_{1t}, \sin\theta_{2t}, 2+\delta-\cos\theta_{1t}-\cos\theta_{2t}).
\label{eq:Bfield}
\end{equation}

Here, $\theta_{lt} = \omega_l t+ \theta_{l0}$ is the phase of the $l$-th drive, and $\omega_1$ and $\omega_2$ are the two driving frequencies. We can assume the ratio between the frequencies is of order 1, then $\omega := \sqrt{\omega_1 \omega_2}$ is the single frequency scale on which the spin-half is driven. $\boldsymbol{\sigma} = (\sigma_x, \sigma_y, \sigma_z)$ is the vector of Pauli matrices. 

Ref.~\cite{crowley2020} investigated the energy pumping power between the two drives, especially for the case $\delta \simeq 0$,
when the time-quasiperiodic system maps to a 2D BHZ model at the phase transition, with a gapless point $(\theta_{1t},\theta_{2t})\simeq (0,0)$ in the 2D (synthetic) BZ. Obviously, the adiabatic condition breaks down in the vicinity of this point,
and the adiabatic energy current given by Eq.~(\ref{eq:adiabatic_current}) should no longer be valid.

To account for the non-adiabatic effects, Ref.~\cite{crowley2020} decomposed the pumping power into two parts: a ``topological'' component and an ``excitation'' component. The topological piece, denoted as $P_{Ts}$ (where the subscript $s$ denotes quantities associated with the driven spin), arises from the spin sampling the Berry curvature of the (synthetic) band on which it stays. The excitation component, denoted as $P_{Es}$, comes from the non-adiabatic excitations when the gap size is comparable to the driving frequencies. To extract the scaling behaviors of these two components, one needs to also consider the complex-conjugated Hamiltonian, $H'$, which can be physically implemented by reversing the chirality of one of the drives. Noting that under complex conjugation, the topological component changes sign, whereas the excitation component remains invariant, Ref.~\cite{crowley2020} was able to write $P_{Es}$ and $P_{Ts}$ as:
\begin{gather}
[P_{Ts}]_{\boldsymbol{\theta_0}} =
    \frac{1}{2} \left( [P_s]_{\boldsymbol{\theta_0}} - [P'_s]_{\boldsymbol{\theta_0}} \right)
\label{eq:PTs} \\
    [P_{Es}]_{\boldsymbol{\theta_0}} =
    \frac{1}{2} \left( [P_s]_{\boldsymbol{\theta_0}} + [P'_s]_{\boldsymbol{\theta_0}} \right),
\label{eq:PEs}
\end{gather}
where $[\cdot]_{\boldsymbol{\theta_0}}$ denotes averaging over the initial phase vector $\boldsymbol{\theta_0}=(\theta_{10},\theta_{20})$, and $P_s$ and $P'_s$ are the pumping powers of the driven spin and its complex-conjugated partner, respectively.

Furthermore, it was also shown that these quantities at short times have universal forms as~\cite{crowley2020}
\begin{gather}
    [\widetilde{P}_{Ts}]_{\boldsymbol{\theta_0}} := 
    \frac{[P_{Ts}]_{\boldsymbol{\theta_0}}}{\omega^2/2\pi} \simeq 
    \frac{1}{1+e^{\alpha x}};
\label{eq:P_Ts} \\
    [\widetilde{P}_{Es}]_{\boldsymbol{\theta_0}} 
    :=\frac{[P_{Es}]_{\boldsymbol{\theta_0}}}{\sqrt{B_0}\omega^{3/2}}
    \simeq C  e^{-\beta x^2},
\label{eq:P_Es}
\end{gather}
where the dimensionless quantity $x= - \delta \sqrt{B_0/\omega}$ characterizes the adiabaticity of the system. 
Here, $\alpha$, $\beta$, and $C$ are dimensionless parameters that can be extracted numerically.
Note that for convenience, we have introduced the rescaled version of the topological and excitation pumping powers, $[\widetilde{P}_{Es}]_{\boldsymbol{\theta_0}}$ and $[\widetilde{P}_{Ts}]_{\boldsymbol{\theta_0}}$, which are dimensionless. 

To gain further intuition, consider first when $|x|$ is large. Indeed, in the adiabatic regime, $[\widetilde{P}_{Ts}]_{\boldsymbol{\theta_0}}$ is effectively 0 (1) for positive (negative) $x$, which is a consequence of the spin uniformly sampling the Berry curvature of the band it is initialized on. Over time, the pumping power will be proportional to the band's Chern number, which is 0 (1) for negative (positive) gap size $\delta$. For the same reason, since there is effectively no transition between the two bands in the adiabatic regime and thus no excited states, $[\widetilde{P}_{Es}]_{\boldsymbol{\theta_0}}$ can be treated as zero.

Near the gapless points, however, the driving frequencies is comparable to the gap size i.e., $|x| \lesssim 1$, and one needs to take the non-adiabatic heating of the spin into account. Indeed, in this regime, $[\widetilde{P}_{Es}]_{\boldsymbol{\theta_0}}$ is Gaussian in $x$, peaking at the gapless point $x=0$. On the other hand, as $x$ increases from $-1$ to 1, $[\widetilde{P}_{Ts}]_{\boldsymbol{\theta_0}}$ smoothly crosses over from 1 to 0 in a Fermi-function-like manner. Note that at the gapless point, $[\widetilde{P}_{Ts}]_{\boldsymbol{\theta_0}}$ takes on the value of $\frac{1}{2}$.

\subsection{Chain of two-level systems under two drives}
In the following, we consider a chain of two-level systems aligned in the $z$-direction, driven by two external drives at mutually irrational frequencies; see Fig.~\ref{fig:model_illustration}(a). The Hamiltonian for such a system can be written as
\begin{equation}
    H = \sum_z \Psi_z^\dagger \mathcal{H}(t) \Psi_z + (\Psi_z^\dagger V \Psi_{z+1} + h. c.),
\label{eq: drivenHam}
\end{equation}
where the sum runs over all sites and $\Psi_z = (\psi_{z,\uparrow},\psi_{z,\downarrow})^T$ 
is the spinor consisting the annihilation operators of
the two orbitals (labeled by $\uparrow$, $\downarrow$)
at site $z$. Similarly, $\Psi_z^\dagger$ is the creation spinor. The onsite, time-dependent potential due to the two external drives has the form
\begin{align}
    &\mathcal{H}(t) =
    -\frac{1}{2} B_0 \Big[ \sin{\theta_{1t}} \sigma_x + \sin{\theta_{2t}} \sigma_y \nonumber \\ &+
    \big(2+\gamma-\cos{\theta_{1t}}-\cos{\theta_{2t}}\big) \sigma_z \Big],
\label{eq: drivenOnsite}
\end{align}
where $\theta_{it}=\omega_i t + \theta_{i0}$, and $\omega_1, \omega_2$ are the two (incommensurate) driving frequencies. Throughout this work, we take the frequency ratio to be the golden ratio, i.e., $\omega_2/\omega_1 = \frac{\sqrt{5}+1}{2} \approx 1.618$. For simplicity, we assume that all sites are initialized with the same phase $(\theta_{01}, \theta_{02})$.
The time-independent hopping term is chosen to be
\begin{equation}
    V = -\frac{B_0}{4} \sigma_z.
\label{eq:hopping}
\end{equation}

To derive the system in synthetic space based on the mapping introduced in Sec.~\ref{sec:quasiperiodic}, we can simply make the substitution, 
$\theta_{it}=\omega_i t + \phi_i \rightarrow k_i$~\cite{GilPRX}, to obtain the two additional Bloch momenta $k_1$, $k_2$ (or $k_x$, $k_y$). 
If we further impose periodic boundary condition along the physical ($z$) direction of the chain, we may introduce the third Bloch momentum $k_z$. 

Thus, we obtain the following 3D Bloch Hamiltonian with inversion symmetry on a cubic lattice, where the lattice spacing is taken to be 1:
\begin{align}
    H(\vec{k}) &= -\frac{1}{2} B_0 \Big[ \big(2+\gamma-\text{cos}(k_x)-\text{cos}(k_y)-\text{cos}(k_z)\big)  \sigma_z \nonumber \\
    &+ \text{sin}(k_x) \sigma_x +  \text{sin}(k_y) \sigma_y \Big].
\label{eq:WSM}
\end{align}

The parameter $\gamma$ controls the phase that the Hamiltonian Eq.~(\ref{eq:WSM}) describes~\cite{RMP}. For $\gamma < -1$ and $\gamma > 1$, it describes a trivial insulator and a topological insulator, respectively. For intermediate values $-1 < \gamma < 1$, Eq.~(\ref{eq:WSM}) describes a WSM phase, with two gapless points, called the Weyl nodes, at $\pm \vec{k_0} = (0,0, \pm k_0)$, where $k_0 = \arccos(\gamma)$. 
Thus, for $-1 < \gamma < 1$, the adiabatic condition for the time-quasiperiodic 1D chain breaks down near $\pm \k_0$. 

One important perspective to understand the WSM is that it can be thought of as a stacking of 2D Chern insulators \cite{RMP}. Indeed, if we fix $k_z$ in Eq.~(\ref{eq:WSM}) and compute the Chern number $C(k_z)$ associated with each of the two-dimensional planes in momentum space, we find that $C(k_z) = 1$ for $|k_z| < |k_0|$, and $C(k_z) = 0$ for $|k_z| > |k_0|$,  i.e., the planes corresponding to fixing $k_z = \pm k_0$ can be regarded as the phase transition points between the 2D nontrivial Chern insulators and the trivial insulators. 

\subsection{Non-adiabatic energy pumping}
In the following, we shall derive the energy pumping power that characterizes the synthetic WSM phase by considering the contribution from each of the 2D (synthetic) Chern insulators (the BHZ model) at fixed $k_z$, based on the perspective described above. In particular, we consider the chain of $N$ sites with periodic boundary conditions at half filling. 
This enables us to apply the known results for non-adiabatic energy pumping in the quasiperiodically driven spin 1/2 introduced in Sec.~\ref{sec:spin-half}.

As discussed in Ref.~\cite{crowley2020}, the non-adiabatic energy pumping can be decomposed into the topological and excitation components (referred to as T and E components hereafter). 
Using the short-time results for the driven single spin in Eqs.~(\ref{eq:P_Ts}) and (\ref{eq:P_Es}), we obtain both components of the phase-averaged energy pumping power (per site) of the driven 1D chain
\begin{align}
 [\widetilde{P}_T]_{\boldsymbol{\theta_0}} &= \frac{2 \pi}{N}\sum_{k_z}[\widetilde{P}_{Ts}]_{\boldsymbol{\theta_0}}(k_z)
 \simeq \int_{-\pi}^\pi dk_z  [\widetilde{P}_{Ts}]_{\boldsymbol{\theta_0}}(k_z) \nonumber \\
 &\simeq 2 k_0
\label{eq:PT} \\
[\widetilde{P}_E]_{\boldsymbol{\theta_0}} &= 
\frac{2 \pi}{N}\sum_{k_z}[\widetilde{P}_{Es}]_{\boldsymbol{\theta_0}}(k_z)
\simeq \int_{-\pi}^\pi dk_z  [\widetilde{P}_{Es}]_{\boldsymbol{\theta_0}}(k_z) \nonumber \\
&\simeq C \sqrt{\frac{\pi}{\beta}} \bigg(\text{erf}\Big(\frac{\gamma+1}{\sqrt{\omega}}\Big) - \text{erf}\Big(\frac{\gamma-1}{\sqrt{\omega}}\Big) \bigg)
\label{eq:PE}
\end{align}
where $k_0 := \arccos{\gamma}$, $\text{erf}(\cdot)$ denotes the Gaussian error function, defined as $\text{erf}(x) = \frac{2}{\sqrt{\pi}} \int_0^x \exp(-t^2) dt$, and the prefactor of 2 arises because of the inversion symmetry of the model.

\begin{figure}
    \centering
    \includegraphics[width=0.49\textwidth]{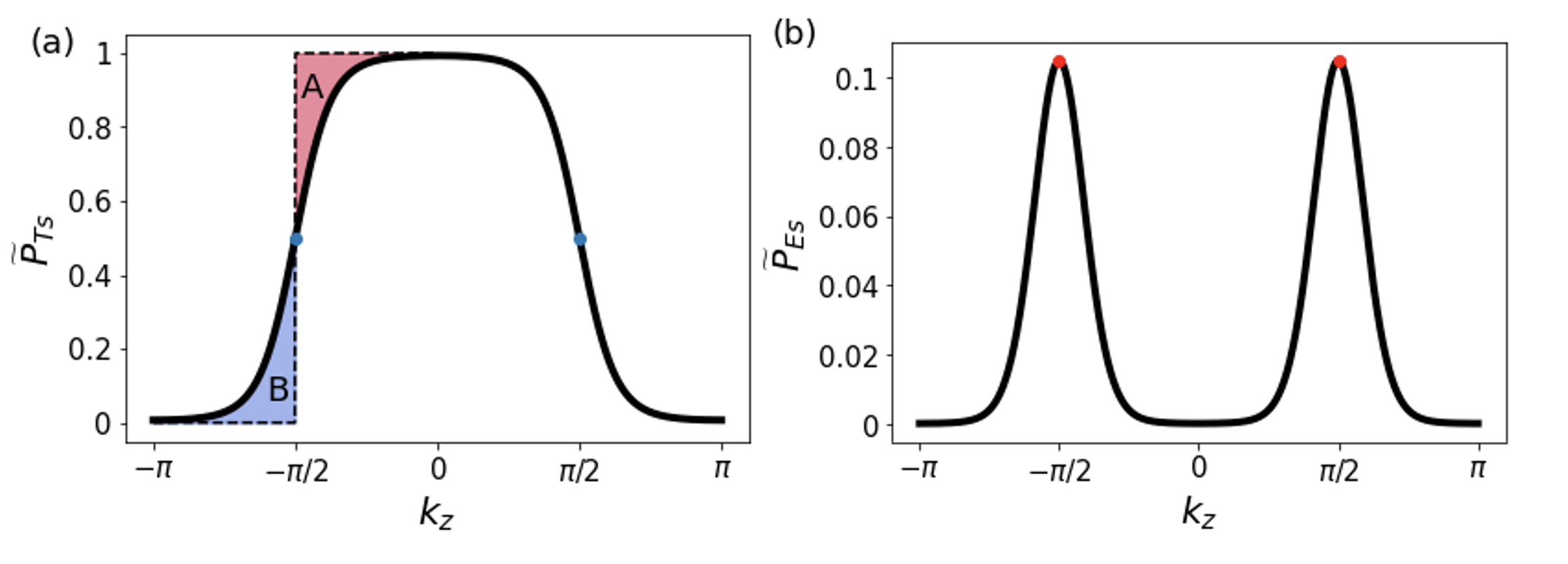}
    \caption{Illustrations of the (a) topological component and (b) excitation component of the pumping power of a synthetic Chern insulator labelled by $k_z$. As illustrated in (a), the ``shortage'' in the topologically nontrivial region (Area A) is filled by the ``surplus'' in the topologically trivial region (Area B), and the pumping power is therefore proportional to the range with nontrivial pumping ($2 \arccos{\gamma}$). The width of the crossover region in (a) and the excitation region in (b) both are of order $\sqrt{\omega/B_0}$. The parameter used is $\gamma = 0 \rightarrow k_0 = \pi/2$.}
    \label{fig:illustration}
\end{figure}

To understand the topological contribution, recall from Sec. \ref{sec:spin-half} that the intergrand of Eq. (\ref{eq:PT}) takes on the value of 0 (1) for sufficiently large (small) $k_z$ and smoothly crosses from 1 to 0 as $k_z$ approaches the Weyl point $k_0$ from below. The crossover region exhibits an approximate symmetry about the Weyl point $\pm k_0$: $[\widetilde{P}_{Ts}]_{\boldsymbol{\theta_0}}(k_z \mp k_0) \simeq 1-[\widetilde{P}_{Ts}]_{\boldsymbol{\theta_0}}(\pm k_0-k_z)$. Therefore, we can fill the ``shortage'' of pumping power in the supposedly topological regime (Area A in Fig. \ref{fig:illustration}(a)) with the ``surplus'' of pumping in the supposedly trivial regime (Area B in Fig. \ref{fig:illustration}(a)). Thus, when computing the topological component, we can effectively treat it as if the system is driven adiabatically: namely, $\widetilde{P}_{Ts} = 1$ for $|k_z| < |k_0|$, and $\widetilde{P}_{Ts} = 0$ for $|k_z| >  |k_0|$. Therefore, the rescaled topological component is equal to the range of $k_z$ associated with nontrivial pumping, or $[\widetilde{P}_T]_{\boldsymbol{\theta_0}} = 2 \arccos{\gamma}$.

To derive the excitation component in Eq.~(\ref{eq:PE}), we have used the fact that $\widetilde{P}_{Es}$ is Gaussian in $\cos{k_z}$, with its peak at $k_z = k_0$ and a width of $\sqrt{\omega/B_0}$. 
Note that in the limit $\omega \rightarrow 0$, $\text{erf}\big(\frac{\gamma\pm 1}{\sqrt{\omega}} \big) \rightarrow \pm 1$, and the above result reduces to $[\widetilde{P}_E]_{\boldsymbol{\theta_0}} =  2C \sqrt{\pi/\beta}$. For this particular system, the parameters extracted by Ref. \cite{crowley2020} are $C = 0.105$ and $\beta = 2.13$, which will be used throughout the rest of this paper. With these parameters, for small $\omega$, we expect $[\widetilde{P}_E]_{\boldsymbol{\theta_0}} \simeq 0.25$.

We remark that our results above are true even beyond the small $\omega$ limit. 
As long as the WSM phase is not too close to a topological insulator or a trivial insulator (more precisely, as long as $\sqrt{\omega/B_0} <1-|\gamma|$), we expect our treatments above to be valid.

\section{Numerical Results \label{sec:numerical}}
\subsection{Periodic boundary condition}
We first present results from simulations of the driven 1D chain under periodic boundary condition, whose Hamiltonian is given by Eq.~(\ref{eq: drivenHam}). 
In particular, we numerically compute the energy current pumped by drive 1, which is given by 
\begin{align}
     J_1(t, \boldsymbol{\theta_0})  &= \omega_1 \Tr\left( \hat{\rho}(t) \frac{\partial H(\boldsymbol{\theta_0})}{\partial \theta_{1t}}  \right)
\label{eq:current}
\end{align}
under the general formalism described in Sec.~\ref{sec:density_matrix}, where the density matrix $\hat{\rho}(t)$ is as defined in Eq.~(\ref{eq:dm_N}). Note that $J_1$ depends on the initial phase $\boldsymbol{\theta_0}$ explicitly.

We simulate the two components of the phase-averaged energy pumping power, $[P_T]_{\boldsymbol{\theta_0}}$ and $[P_E]_{\boldsymbol{\theta_0}}$, in the following way. 
We first randomly choose a point $\boldsymbol{\theta_0} = (\theta_{01}, \theta_{02}) \in (0, 2\pi)^2$ on the torus as our initial phase vector of the two drives and fix the time-dependent Hamiltonian $H(t, \boldsymbol{\theta_0})$ given by Eq.~(\ref{eq: drivenHam}).
Next, we initialize the system in its ground state at half-filling, described by the density matrix $\hat{\rho}(t=0)$. Then we perform time evolution on the system using $H(t, \boldsymbol{\theta_0})$. At each time, we compute the expectation value of the energy current according to Eq.~(\ref{eq:current}). Upon integrating over time, we obtain the total work done by drive 1.

Starting from the ground state of the complex-conjugated Hamiltonian $H'(t, \boldsymbol{\theta_0})$, we follow the same steps and compute the energy current $J_1'(t, \boldsymbol{\theta_0})$ in the complex-conjugated system. 
Note that this conjugation can be physically implemented by reversing the chirality of one of the polarizing drives \cite{crowley2020}. Similar to the driven spin-half system reviewed in Sec. \ref{sec:spin-half}, we can extract the (rescaled) topological (T) and excitation (E) components as:
\begin{align}
 [\widetilde{P}_T]_{\boldsymbol{\theta_0}}(t) = \frac{\pi}{N} \frac{1}{\omega^2/2\pi}  \Big[ J_1(t, \boldsymbol{\theta_0}) - {J_1'}(t, \boldsymbol{\theta_0}) \Big]_{\boldsymbol{\theta_0}}; 
\label{eq:PT_numerical} \\
[\widetilde{P}_E]_{\boldsymbol{\theta_0}}(t) = \frac{\pi}{N} \frac{1}{\sqrt{B_0} \omega^{3/2}} \Big[ J_1(t, \boldsymbol{\theta_0}) + {J_1'}(t, \boldsymbol{\theta_0}) \Big]_{\boldsymbol{\theta_0}}.
\label{eq:PE_numerical}
\end{align}

\begin{figure}
    \centering
    \includegraphics[width=0.42\textwidth]{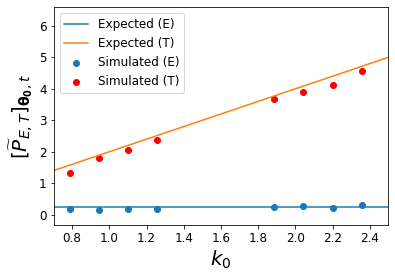}
    \caption{A plot of time- and phase-averaged, rescaled pumping powers, $[\widetilde{P}_T]_{\boldsymbol{\theta_0}, t}$ and $[\widetilde{P}_E]_{\boldsymbol{\theta_0}, t}$, vs. the position of the Weyl node $k_0 = \arccos{\gamma}$. The expected behaviors for the topological and excitation components, shown in orange and blue, are linear and constant in $k_0$ respectively. The parameter used are $\omega/B_0 = 0.1, N = 40$.}
    \label{fig:vsGamma}
\end{figure}

\begin{figure}
    \centering
    \includegraphics[width=0.42\textwidth]{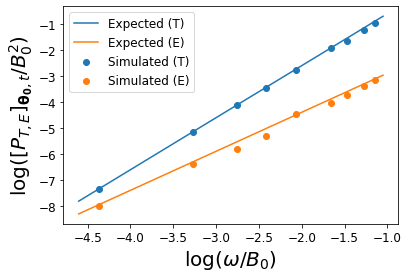}
    \caption{A log-log plot of time- and phase-averaged pumping powers vs. the frequency scale $\omega/B_0$. To retain the frequency dependence, the powers are \textit{not} rescaled. The topological and excitation components are expected to be linear with slopes 2 and 3/2 (shown by the blue and orange lines respectively.) For illustration purposes, the expected and simulated topological piece has been shifted up by $\log(2 \pi)$. The parameter used for all data points is $\gamma = -0.454$. The number of sites $N$ is 120 for $\omega/B_0 = 0.01$, 80 for $\omega/B_0 = 0.03$, and 40 for the rest of the frequencies.}
    \label{fig:vsOmega}
\end{figure}
We simulate the driven chain and its complex-conjugated partner up to time $T_0 := 1.72 \omega^{-3/2}$.
Furthermore, we compute the time-averaged pumping power by dividing the total work done by drive 1 by the total evolution time $T_0$:
\begin{equation}
    [\widetilde{P}_{T,E}]_{\boldsymbol{\theta_0}, t}  = \frac{1}{T_0}\int_0^{T_0} dt' [\widetilde{P}_{T,E}]_{\boldsymbol{\theta_0}}(t').
\end{equation}

In Fig.~\ref{fig:vsGamma}, we show the simulation results of the time-averaged pumping power as a function of $k_0 := \arccos{\gamma}$.
The expected behaviors, given by Eqs.~(\ref{eq:PT}) and (\ref{eq:PE}), are shown in solid lines in Fig. \ref{fig:vsGamma} for comparison. Here the parameters used are $\omega/B_0 = 0.1, N = 40$. For each $k_0$, we average over 400 initial phases. 

We next numerically investigate the frequency dependence. While the rescaled pumping powers do not explicitly depend on the frequency $\omega$, the \textit{unrescaled} topological and excitation components are proportional to $\omega^2$ and $\sqrt{B_0}\omega^{3/2}$, respectively. Therefore, in a log-log plot, we expect the T and E components to be lines with slopes 2 and 3/2, and intercepts $\log{(2 k_0)}$ and $\log{\widetilde{P_E}}$, respectively. In Fig. \ref{fig:vsOmega}, we show the simulated and expected results. The parameter used are $\gamma = -0.454$ and 400 initial phases. The number of sites used are $N = 120$ for $\omega/B_0 = 0.01$, $N = 80$ for $\omega/B_0 = 0.03$, and $N = 40$ for the rest of the frequencies. We remark that there has to be more sites for smaller frequencies because the width of the ``nonadiabatic region'' scales as $\sqrt{\omega/B_0}$, so one needs a denser sampling of the physical Brillouin zone to fully capture the Fermi-function-like and Gaussian functions of the T and E contributions near the transition point.

\subsection{Open boundary condition \label{sec:bc}}
After showing results for the chain under periodic boundary condition, we next discuss the case with open boundary condition along the $z$-direction, which is more naturally prepared in the laboratory.
\begin{figure}
    \centering
    \includegraphics[width=0.4\textwidth]{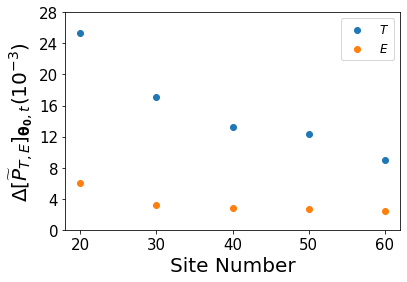}
    \caption{A plot of the pumping power difference between a driven chain and a driven ring, averaged over time and the \textit{same initial phases}. As the number of sites increases, the finite-size effect diminishes. Note that even for $N = 20$, the difference is considerably small, meaning that one can replace a driven ring with a driven ring in a realistic experimental setting. The other parameters used are $\omega/B_0 = 0.1, \gamma = -0.309$.}
\label{fig:BC}
\end{figure}

In Fig.~\ref{fig:BC}, we plot the difference in the pumping powers (per site) of a driven chain (open boundary condition) and a driven ring (periodic boundary condition) vs. the number of sites $N$. To better control the variables, for each site number $N$, we simulated the pumping powers of a driven ring and a driven chain with the same 200 initial phases. We see that the finite-size effect indeed vanishes as the site number increases. Moreover, note that the difference is small even for $N = 20$, which means that practically, one can experimentally realize the synthetic WSM phase with desired pumping powers in a driven chain.

\subsection{Spatial disorder}
Thus far, we have assumed the chain is ideal in that the onsite driving potentials are the same for all sites.
In reality, however, the chain is likely to contain some type of spatial disorder. 
For example, each site may feel a slightly different time-dependent potential. 
Therefore, in the following, we introduce real-space onsite disorder to the driven chain
to study the disorder effects on the energy pumping phenomena in the synthetic WSM.
In particular, for each of the $N$ sites, 
we add a disorder term to the overall driving amplitude by modifying the onsite driving term in Eq. (\ref{eq: drivenOnsite}) to 
\begin{align}
    \mathcal{H}_i(t) &=
    -\frac{1}{2} (1+\epsilon_i) B_0 \left[\sin\theta_{1t}\sigma_x + \sin\theta_{2t}\sigma_y \right. \nonumber  \\ 
    &+ \left. \left(2+\gamma-\cos\theta_{1t}-\cos\theta_{2t}\right) \sigma_z \right]
\label{eq:disordered}
\end{align}
where $i$, running from 1 to $N$, is the site index, and $\epsilon_i$ is the drawn from a uniform distribution on the interval $[-\eta/2, \eta/2]$. The parameter $\eta$ characterizes the strength of the onsite disorder; in the following, we will consider three values of $\eta$, $\eta = 0.01, 0.1, 0.5$, which correspond to the cases where the onsite disorder strength is much less than, less than, and comparable to the driving amplitude, respectively.

\begin{figure}
    \centering
    \includegraphics[width=0.48\textwidth]{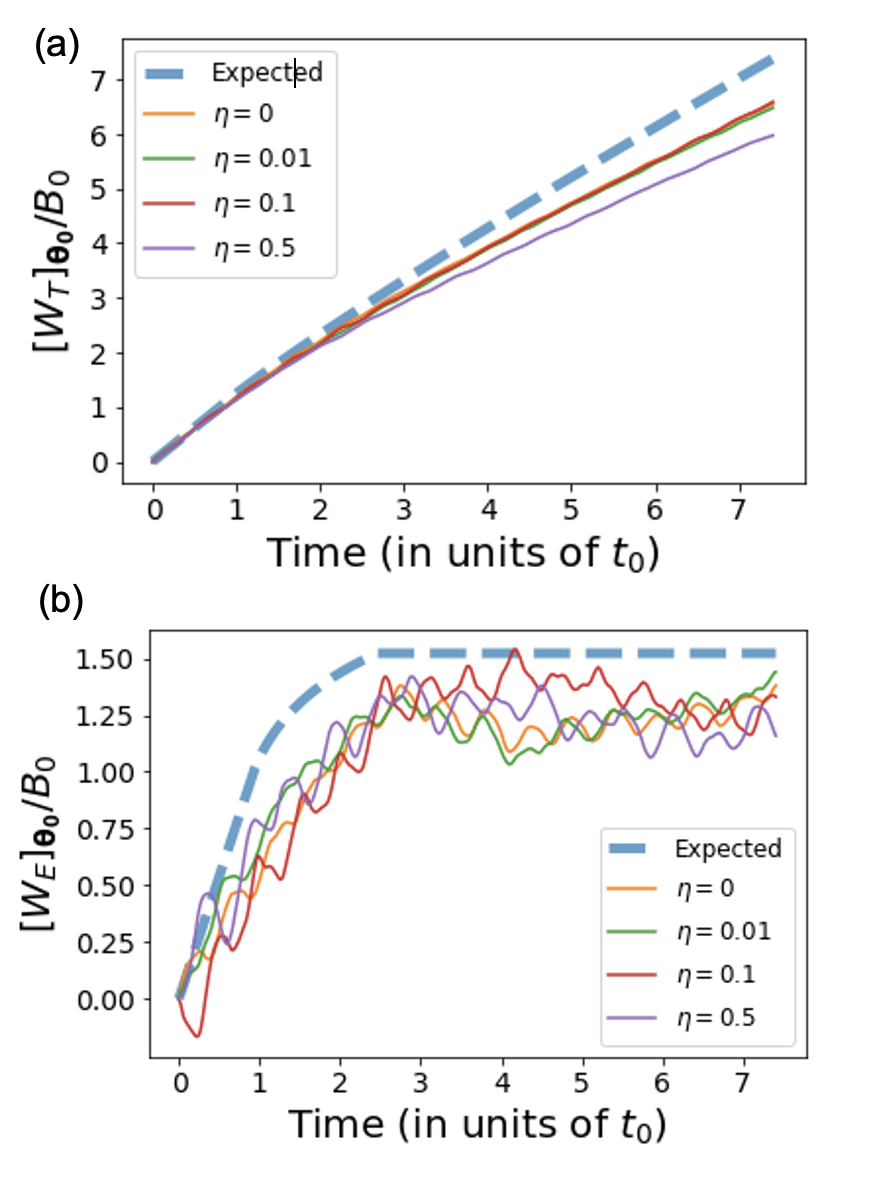}
    \caption{The (a) topological and (b) excitation components of the energy pumped as a function of time. The expected behaviors of the two components, given by Eq.~(\ref{eq:longTexpected}), are shown in dashed lines. The simulated pumping behaviors in the presence of disorder are given by Eq.~(\ref{eq:energy_pumped}) and shown in solid lines. The parameter $\eta$ characterizes the onsite disorder strength. We numerically confirm that both components are surprisingly robust against spatial disorder, and the simulated results are in good agreement with the expected ones.
    All plots are not rescaled by $\omega$ or $B_0$, and in the presence of disorder, they are \textit{without} disorder averaging. The parameter used are $\omega/B_0 = 0.1, N = 40, \gamma =  -0.454$, $t_0 = 5.41 \sqrt{B_0/\omega^3}$ (for a discussion of the unlocking time, see Sec. \ref{sec:discussions}).}
\label{fig:longtime}
\end{figure}

In Fig.~\ref{fig:longtime}, we show the (un-rescaled) simulated energy pumping behaviors of the topological and excitation components as a function of time in solid lines. The quantity shown is defined as
\begin{equation}
[W_{T,E}]_{\boldsymbol{\theta_0}}(t) = \int_0^t dt' [P_{T,E}]_{\boldsymbol{\theta_0}}(t').
\label{eq:energy_pumped}
\end{equation}

Importantly, we remark that the pumping behaviors in Fig. \ref{fig:longtime} are \textit{without} spatial disorder averaging (they are, however, averaged over the initial phase vectors as before). It is clear that both components of the pumping power are surprisingly robust against onsite disorder -- even for $\eta = 0.5$, in which case the disorder strength is comparable to the driving strength. 

\section{Unlocking time and energy pumping \label{sec:discussions}}
In this section, we first give a more detailed analysis of the timescale at which an average spin-1/2 decouples from the driving magnetic field (hereafter referred to as the unlocking time $t_u$). Understanding how $t_u$, the most important timescale, scales with various quantities of the system in turn allows us to derive the expected energy pumping behaviors of the synthetic WSM.

Let us begin by considering a spin-1/2 introduced in Sec.~\ref{sec:spin-half}, where the two driving frequencies are of scale $\omega$, the gap size is $B_0 \delta$, and the magnetic field is given by Eq.~(\ref{eq:Bfield}). After some characteristic time $t_u$, the dynamics of the driven spin unlocks from the driving field, and the pumping power effectively becomes zero in a statistical sense. More precisely, at late times $t \gtrsim t_u$, the populations in the ground and excited states become equal. The topological component of pumping vanishes because the Chern numbers of the ground and excited bands sum up to zero, whereas the excitation component vanishes because an average spin in the ensemble ceases to absorb energy~\cite{crowley2020}. Using the Landau-Zener formula~\cite{Landau, Zener}, we have the probability that a spin in the ground state would enter the excited state scales as
\begin{equation}
    p_\text{exc} \sim  \exp{\left(-(B_0 \delta)^2 \Big/ B_0 \sqrt{\omega_1^2 + \omega_2^2}\right)}
    \label{eq:p_exc},
\end{equation}
where $\delta$ is the gap size in~\ref{eq:Bfield}. The above equation states that the spin unlocks from the magnetic field when the square of the gap is comparable to the rate of change of the field. We note in passing that since the excitation component of the pumping power $\widetilde{P}_{Es}$ is proportional to the excited population and thus to $p_\text{exc}$, Eq.~(\ref{eq:p_exc}) accounts for the Gaussian shape of $\widetilde{P}_{Es}$. Eq.~(\ref{eq:p_exc}) also immediately implies that the timescale at which the spin unlocks will take the form~\cite{GilPRX}:
\begin{equation}
    t_u \sim \frac{1}{p_{exc}} = t_0 \exp{\left((B_0 \delta)^2 \Big/ \sqrt{\omega_1^2 + \omega_2^2}\right)},
    \label{eq:tu}
\end{equation}
where $t_0$ is the unlock time at $\delta = 0$. 
Ref.~\cite{crowley2020} was able to write down $t_0 \sim \sqrt{B_0/\omega^3}$. Here, we take the argument a step further and obtain a good prefactor for $t_0$. By combining the explicit form of the magnetic field, Eq.~(\ref{eq:Bfield}), with the Landau-Zener formula, we can write down the condition of unlock as $B_0^2 (\theta_{1t}^2 + \theta_{2t}^2) \lesssim B_0 \sqrt{\omega_1^2 + \omega_2^2}$, or $|\boldsymbol{\theta}(t)|^2 \lesssim \sqrt{\omega_1^2 + \omega_2^2}/B_0$, where $\boldsymbol{\theta}(t) = (\theta_{1t}, \theta_{2t})$ is a vector containing the phases of the two drives. It is thus helpful to define a circular ``excitation region'' in the $\theta$-space with a center at $(0, 0)$ and a radius of $\theta^*$, such that a spin unlocks from the driving field when its trajectory in phase space first enters this region~\cite{crowley2020}. For our particular choice of frequency ratio $\omega_2/\omega_1 = \frac{\sqrt{5}+1}{2}$, we expect $\theta^* \approx 1.22 \sqrt{\omega/B_0}$. 

\begin{figure}
    \centering
    \includegraphics[width=0.49\textwidth]{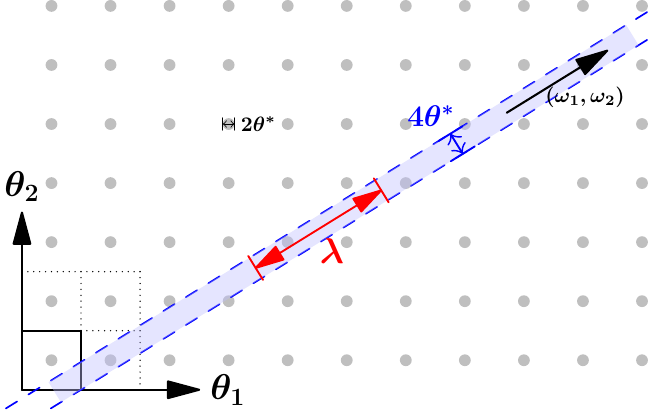}
    \caption{An illustration for a typical spin trajectory in $\theta$-space. Note that due to periodic boundary conditions, we have extended the phase-space trajectory from a 2-torus to the full 2D plane. In the figure, the squares denote (synthetic) Brillouin zones, and the gray circles denote excitation regions, which have diameters $2 \theta^*$. The arrow in black denotes the spin's trajectory along the $(\omega_1, \omega_2)$ direction. Note that for a given trajectory, the only possible excitation regions that the trajectory can enter will be enclosed inside a stripe of width $4\theta^*$, broadened symmetrically from the trajectory, as indicated in light blue.
    The mean free path of the trajectory $\lambda$ is indicated in red, such that the area $4\theta^*\lambda$ contains one excitation region in average.  }
\label{fig:tu}
\end{figure}
To obtain an estimation of $t_0$, we can use the idea of ``mean free path" of spin trajectories, 
which is the average distance that a spin travels in the (synthetic) BZ before entering an excitation region.
 Note first that since $\dot{\theta}_{lt} = \omega_l$, the spins move along the $(\omega_1, \omega_2)$ direction in $\theta$-space. Furthermore, since $\theta_1, \theta_2$ are $2\pi$-periodic variables, we may extend the phase's trajectory from a 2-torus to the entire 2D real plane; see Fig.~\ref{fig:tu}. In this figure, each square denotes a Brillouin zone, the circle at the square center denotes the excitation region, and the black arrow denotes the spin trajectory. 
 
 For a given trajectory, the only possible excitation regions that the trajectory can enter will be enclosed inside a stripe of width $4\theta^*$, broadened symmetrically from the trajectory. Thus, the mean free path $\lambda$, should be defined such that the area $4\theta^*\lambda$
 contains one excitation region in average. This leads to $\lambda = (4\theta n)^{-1}$, where $\lambda$ is the mean free path, and $n$ is the number of excitation region per unit area. Here, $n = (4 \pi^2)^{-1}$ because there is one excitation region per BZ, which has area $(2 \pi)^2$. 
Therefore, $\lambda \approx \pi^2/\theta^*$, from which we obtain the estimated $t_0$ as:
\begin{equation}
    t_0 = \frac{\lambda}{\sqrt{\omega_1^2 + \omega_2^2}} \approx 5.41 \sqrt{B_0/\omega^3}.
\end{equation}

Next we turn to the expected energy conversion behavior of a synthetic WSM. Recall from Sec.~\ref{sec:pumpingpower} that we can view a synthetic WSM as a stack of synthetic Chern insulators i.e., as a stack of driven spin-1/2's. These driven spins are labelled by $k_z$ and have gap sizes $\delta = \gamma -\cos{k_z}$. For $t < t_0$, we expect that all of the spins remain locked to the driving field. As the evolution time goes past $t_0$, however, spins with $B_0 \delta \simeq 0$, i.e. spins labelled by $k_z \simeq k_0$, start to unlock. More precisely, From Eq.~(\ref{eq:tu}), we expect that at time $t$, spins with gap size $B_0 \delta \lesssim \sqrt{B_0 \omega \log(t/t_0)}$ cease to pump energy. From this, we can write down:
\begin{equation}
    [\widetilde{P}_{T,E}]_{\boldsymbol{\theta_0}}(t') = \int_{-\pi}^{\pi} dk_z [\widetilde{P}_{Ts,Es}]_{\boldsymbol{\theta_0}}(t';k_z),
    \label{eq:longTexpected}
\end{equation}
where
\begin{gather}
    [\widetilde{P}_{Ts}]_{\boldsymbol{\theta_0}}(t';k_z) \simeq \frac{1}{1+e^{\alpha x}}\Theta(t_u(\gamma - \cos{k_z}) - t'),\\
    [\widetilde{P}_{Es}]_{\boldsymbol{\theta_0}}(t';k_z) \simeq Ce^{-\beta x^2}\Theta(t_u(\gamma - \cos{k_z}) - t'),
\end{gather}
are the pumping powers for the T and E components of a driven spin labelled by $k_z$, as a function of time. Here $x := (\cos{k_z} - \gamma)\sqrt{B_0/\omega}$ is the adiabaticity parameter as before, and $\Theta(t) := \mathbf{1}_{t>0}$ is the Heaviside step function.

From the above results, we can understand what to expect for the pumping power of a synthetic WSM phase. Prior to $t_0$, since all spins maintain their fidelity, the T and E components should both be ideal, given by Eqs.~(\ref{eq:PT}) and~(\ref{eq:PE}) respectively. The components' long-time behaviors, however, differ considerably. Consider first the T component. We expect that this component remains nontrivial for exponentially long times, because spins with a large gap (relative to the driving frequency) maintain their fidelity -- and therefore contribute to pumping -- for exponentially long times. On the other hand, since $[P_E]_{\boldsymbol{\theta_0}}$ is nontrivial for $\delta \lesssim \sqrt{\omega/B_0}$ only, we expect from Eq.~(\ref{eq:tu}) that after $t_u \approx \text{e} t_0$, the excitation region will fully unlock, and the E component will vanish. This is indeed what we see from the simulation results Fig.~\ref{fig:longtime}(b) -- after $t_u \sim 2.7 t_0$, the excitation component effectively becomes flat, corresponding to a fully unlocked excitation region. Importantly, we note that this is true for all disorder strengths. From the analysis above, we can numerically compute the expected T and E components of energy using Eq.~(\ref{eq:energy_pumped}); see the dashed lines in the two panels in Fig.~(\ref{fig:longtime}). Note that while we expect $[P_T]_{\vec{\theta_0}} \rightarrow 0$ as $t/t_0 \rightarrow \infty$, our finite-time simulations is not able to capture this behavior, as some spins pump for exponentially long times. It is clear from the concave shape of $W_T(t)$, however, that the topological pumping power decreases as time goes due to more spins unlocking.

\section{Experimental realization}
We propose a concrete experimental setup to realize our 1D model based on ultracold atoms in optical lattices, which appear as promising candidates for simulating intriguing phenomena and phases~\cite{goldman2016topological}, such as the topological Thouless pumping~\cite{nakajima2016topological}, and the 3D topological insulator with axion electrodynamics~\cite{Bermudez2010}. Here, we propose to use fermionic atomic gases (e.g. \ce{^{6}Li}, \ce{^{40}K} atoms)
in optical lattices to engineer a gauge transformed version of the quasiperiodically driven 1D chain proposed previously. The setup
is inspired by the similar one proposed in Ref.~\cite{YangPRB}.

By introducing a gauge tranformation $\widetilde{\Psi}_{z}=U_{z}\Psi_{z}$
with some $z$-dependent unitary matrix $U_{z}$, we can rewrite the Hamiltonian in Eq.~(\ref{eq: drivenHam}) as
\begin{equation}
H=\sum_{z}\widetilde{\Psi}_{z}\widetilde{\mathcal{H}}_{z}(t)\widetilde{\Psi}_{z}+(\widetilde{\Psi}_{z}^{\dagger}U_{z}VU_{z+1}^{\dagger}\widetilde{\Psi}_{z+1}+h.c.),
\end{equation}
where $\mathcal{\widetilde{H}}_{z}(t)=U_{z}\mathcal{H}(t)U_{z}^{\dagger}$.

If we choose
\begin{equation}
U_{z}=\left(\begin{array}{cc}
1 & 0\\
0 & e^{i\pi z}
\end{array}\right),
\end{equation}
the hopping term becomes $U_{z}VU_{z+1}^{\dagger}=-B_{0}\sigma_{0}/4$ which is spin-independent, and
the onsite Hamiltonian becomes
\begin{align}
\widetilde{\mathcal{H}}_{z}(t)&=-\frac{1}{2}B_{0}\left[(-1)^{z}\left(\sin\theta_{1t}\sigma_{x}+\sin\theta_{2t}\sigma_{y}\right)\right. \nonumber\\
& \left. +(2+\gamma-\cos\theta_{1t}-\cos\theta_{2t})\sigma_{z}\right].
\label{eq:gauge_transformed_ham}
\end{align}
Note that here $z\in \mathbb{Z}$ is the site index of the 1D chain, and this term corresponds to a magnetic field in the $xy$ plane alternating between even and odd sites. 

This Hamiltonian can be realized with ultra cold atoms, such as $\ce{^{6}Li}$,
trapped in an optical lattice. Each atom can be described by a two-level
system, due to the hyperfine ground-state manifold with total angular
momentum $F=1/2$. The optical lattice potential $V$ projected to
this ground state manifold can be written as \cite{goldman2016topological,Deutsch1998}
\begin{align}
V(\boldsymbol{r})=V_{0}(\boldsymbol{r})+\boldsymbol{B}_{{\rm eff}}(\boldsymbol{r})\cdot\boldsymbol{\sigma}
+ g \boldsymbol{B_{\text{ext}}}\cdot\boldsymbol{\sigma},
\end{align}
where the Pauli matrices $\boldsymbol{\sigma}=(\sigma_{x},\sigma_{y},\sigma_{z})$
act on this two dimensional ground state manifold. Here the first two terms 
\begin{gather}
V_{0}(\boldsymbol{r})=u_{s}(\widetilde{\boldsymbol{E}}^{*}(\boldsymbol{r})\cdot\widetilde{\boldsymbol{E}}(\boldsymbol{r}))\\
\boldsymbol{B}_{{\rm eff}}(\boldsymbol{r})=iu_{v}(\widetilde{\boldsymbol{E}}^{*}(\boldsymbol{r})\times\widetilde{\boldsymbol{E}}(\boldsymbol{r}))
\end{gather}
are determined by the complex electric field $\widetilde{\boldsymbol{E}}$,
whose components are defined as $\widetilde{E}_{j}=E_{j}\cos(\phi_{j})$
with $j=x,y,z$ of an electric field $\boldsymbol{E}(t)=\sum_{j}E_{j}\cos(\phi_{j}-\omega t)\boldsymbol{e}_{j}$,
where $\boldsymbol{e}_{x,y,z}$ are unit vectors in $x,y,z$ directions.
$\boldsymbol{B}_{{\rm ext}}$ is the external magnetic field applied,
while $g$ is gyromagnetic ratio.

Let us choose
\begin{equation}
\widetilde{\boldsymbol{E}}(\boldsymbol{r})=\left(\cos k_{L}y, \cos k_{L}x, i[\epsilon_{1}\cos k_{L}x+\epsilon_{2}\cos k_{L}y]\right),
\end{equation}
where $k_{L}$ is the wave vector of the laser. This gives
\begin{align}
V_{0}(\boldsymbol{r}) & =u_{s}\left[(1+\epsilon_{1}^{2})\cos^{2}(k_{L}x)+(1+\epsilon_{2}^{2})\cos^{2}(k_{L}y)\right.\nonumber \\
 & +\left.2\epsilon_{1}\epsilon_{2}\cos(k_{L}x)\cos(k_{L}y)\right]
\end{align}
and the effective magnetic field
\begin{align}
\boldsymbol{B}_{{\rm eff}}(\boldsymbol{r}) & =-2u_{v}\cos(k_{L}x)\left[\epsilon_{1}\cos k_{L}x+\epsilon_{2}\cos k_{L}y\right]\boldsymbol{e}_{x}\nonumber \\
 & +2u_{v}\cos(k_{L}y)\left[\epsilon_{1}\cos k_{L}x+\epsilon_{2}\cos k_{L}y\right]\boldsymbol{e}_{y}.
\end{align}
With $u_{s}<0$, and $|\epsilon_{1,2}|$, $|u_{v}\epsilon_{1,2}| \ll 1$,
the atoms will be trapped at the local minima of $V_{0}$ at $(x,y)=(n_{x},n_{y})\lambda_{L}/2,$
with $n_{x},n_{y}\in\mathbb{Z},$ and $\lambda_{L}=2\pi/k_{L}$ is
the laser wave length. The effective magnetic field at the potential
minima becomes
\begin{align}
\boldsymbol{B}_{{\rm eff}}(n_{x},n_{y}) & =-2u_{v}[\epsilon_{1}+(-1)^{n_{x}+n_{y}}\epsilon_{2}]\boldsymbol{e}_{x}\nonumber \\
 & +2u_{v}[\epsilon_{2}+(-1)^{n_{x}+n_{y}}\epsilon_{1}]\boldsymbol{e}_{y}.
\end{align}

By choosing $g\boldsymbol{B}_{{\rm ext}}=(2u_{v}\epsilon_{1},-2u_{v}\epsilon_{2},gB_{z})$,
the constant term in $\boldsymbol{B}_{{\rm eff}}$ can be canceled
out, and we obtain optical lattice potential on the ground state manifold as
\begin{equation}
V(n_{x},n_{y})=u_v(-1)^{n_x + n_y}(-\epsilon_2\sigma_x + \epsilon_1\sigma_y)+gB_z\sigma_z.
\label{eq:optical_potential}
\end{equation}

If we apply an additional confining potential in $y$ and $z$ direction, we can obtain a 1D optical lattice
along $x$ direction with $n_x\in\mathbb{Z}$ and $n_y=0$, with an alternating magnetic field in the $xy$ plane between even and odd sites,
as in Eq.~(\ref{eq:gauge_transformed_ham}). 
We further require $\epsilon_1\sim\sin\theta_{2t}$, $\epsilon_2\sim \sin\theta_{1t}$, and $B_z\sim(2+\gamma-\cos\theta_{1t}-\cos\theta_{2t})$ are time dependent, oscillating at frequencies $\omega_1$, $\omega_2$.
The spin-independent hopping term $-B_0\sigma_0/4$ is given by the overlap of the Wannier
functions centered at neighboring minima of the potential~\cite{goldman2016topological}. 
Thus, we have shown all terms in the gauge transformed Hamiltonian in Eq.~(\ref{eq:gauge_transformed_ham}) can be realized using 
ultracold atoms such as $\ce{^{6}Li}$ in optical lattices. 

With such an experimental setup based on ultra cold $\ce{^{6}Li}$ fermions, with red-detuned laser of wave length around $1000$nm,  the energy scale of the effective Zeeman field due to optical potential can be of order $10$ to $100$kHz, which also requires the external magnetic field at the scale of $10$mG (milliGauss). The driving frequencies $\omega_{1,2}$ should then be $\lesssim 10$kHz. With these parameters, the unlocking time $t_0$ can be of order $1$ms. The topological component of the energy pumping power per site, at short time ($t<t_0$), can be of order $100\mathrm{kHz}/\mathrm{ms}$.

\section{Conclusion\label{sec:conclusions}}
 In this work, we proposed a potential quantum device, 
 a 1D chain quasiperiodically driven by two frequencies, which maps to a synthetic Weyl semimetal. 
 We demonstrate that this device exhibits robust and universal energy pumping/flow that scales linearly in system size, and is able to operate in a relatively large parameter regime without the assumptions of adiabaticity.
In other words, our proposed system goes beyond the ones proposed in 
previous works, which are either extended but gapped in synthetic dimensions~\cite{YangPRB}, or gapless but only of zero dimension (single spin)~\cite{crowley2020}.

In particular, as we showed both analytically and numerically, 
the energy pumping power has simple and universal scaling forms, given in Eqs.~(\ref{eq:PT}) and (\ref{eq:PE}) at short times, and in Eq.~(\ref{eq:longTexpected}) at intermediate/long times.
These behaviors are robust even in the presence of reasonable amount of spatial disorder.

Regarding physical realizations, we proposed an experimental setup based on
ultracold atoms such as $\ce{^{6}Li}$ in optical lattices. 

Finally, we want to comment on the possible interaction effects resulting in heating, 
which also absorbs energies from the quasiperiodic drives, in addition to the nonadiabatic excitations.
It is known that for smooth driving protocols (as the harmonic drives used in our model),
when the interactions are local,
there exists a slow heating regime when the driving time is less than typical time $t_*$~\cite{Else2020}, which is bounded for any $\epsilon>0$, as
\begin{equation}
    t_* \geq \frac{C'}{J}\exp\left[C\left(\frac{\omega}{J}\right)^{1/(m+\epsilon)}\right],
\end{equation}
where $\omega$ is the norm of the frequency vector, $J$ is the local interaction energy scale, $m$ is the number of external drives, and $C,C'$ are dimensionless constants depending on the number-theoretic properties of the ratio between the frequencies.
Because of this, the universal energy pumping behavior discussed in this work for noninteracting system should be detected 
in a short time window before reaching $t_*$ in realistic systems (such as in our proposed cold-atom setup) where interactions do occur.
Since our results do not have the restriction of adiabaticity, which may limit the magnitude of $\omega$, we expect to be able to host a relatively large $t_*$, making the universal energy pumping behavior described in this work more accessible than the simple topological energy pumping 
discussed in the previous works~\cite{GilPRX,YangPRB} which do require small frequency compared to the driving amplitude. 

Apart from potential applications, finding the corresponding signatures in synthetic systems for nontrivial phenomena in physical
systems is of importance on its own. We would like to point out two possible future directions beyond the current work. 
First, the nodal-line semimetals are another type of gapless systems with sets of 1D gapless points in Brillouin zone. It has been
also shown that these systems can be tuned into Weyl semimetals with circularly polarized light \cite{Yan2016}.
It would be intersting to see the synthetic analog of thses systems and their nonadiabatic energy pumping phenomena. 
Second, as is known that the Weyl semimetal has topologically protected Fermi arcs on its boundaries. 
Whether or not the physics of Fermi arcs can result in any intriguiing quantum dynamics of the quasiperiodic systems introduced
in this work, should be addressed in the future. 

\acknowledgments
We acknowledge support from the Institute of Quantum Information and Matter, an NSF Frontier center. G.R. is also grateful for support from the
Simons Foundation and the DARPA DRINQS program. Z.Q. is grateful for support from Caltech's Student-Faculty program and the Victor Neher Fellowship. Y.P. acknowledges support from the startup fund from California State University, Northridge. Numerical calculations have been performed using QuTip~\cite{qutip, qutip2}.  This
work was performed in part at Aspen Center for Physics,
which is supported by National Science Foundation grant
PHY-1607611. 

%\bibliography{main}
%
\end{document}